\documentclass[twocolumn,prb,superscriptaddress,aps,10pt]{revtex4-1} 

\usepackage{amsmath}  
\usepackage{amsfonts} 
\usepackage{graphicx} 
\usepackage{hyperref}
\usepackage{enumitem}
\usepackage{minibox}
\usepackage{color}
\usepackage{subcaption}

\begin{document}

\title{Bridging Physics and Biology Teaching through Modeling}

\author{Anne-Marie Hoskinson}
\email{Author to whom correspondence should be addressed (biology): hoskin29@msu.edu}
\affiliation{Department of Plant Biology, Michigan State University, East Lansing, MI 48824, USA}
\author{Brian A. Couch}
\affiliation{Department of Molecular, Cellular, and Developmental Biology, University of Colorado Boulder, Boulder, CO 80309, USA}
\author{Benjamin M. Zwickl}
\affiliation{School of Physics and Astronomy, Rochester Institute of Technology, Rochester, NY 14623, USA}
\author{Kathleen A. Hinko}
\affiliation{Department of Physics and JILA NSF Physics Frontier Center for Atomic, Molecular and Optical Physics, University of Colorado Boulder, CO 80309, USA}
\author{Marcos D. Caballero}
\email{Author to whom correspondence should be addressed (physics): caballero@pa.msu.edu} 
\affiliation{Department of Physics and Astronomy, Michigan State University, East Lansing, MI 48824, USA}

\date{\today}

\begin{abstract}
As the frontiers of biology become increasingly interdisciplinary, the physics education community has engaged in ongoing efforts to make physics classes more relevant to life sciences majors. These efforts are complicated by the many apparent differences between these fields, including the types of systems that each studies, the behavior of those systems, the kinds of measurements that each makes, and the role of mathematics in each field. Nonetheless, physics and biology are both sciences that rely on observations and measurements to construct models of the natural world. In the present theoretical article, we propose that efforts to bridge the teaching of these two disciplines must emphasize shared scientific practices, particularly scientific modeling. We define modeling using language common to both disciplines and highlight how an understanding of the modeling process can help reconcile apparent differences between the teaching of physics and biology. We elaborate how models can be used for explanatory, predictive, and functional purposes and present common models from each discipline demonstrating key modeling principles.  By framing interdisciplinary teaching in the context of modeling, we aim to bridge physics and biology teaching and to equip students with modeling competencies applicable across any scientific discipline.     
\end{abstract}

\maketitle 

\section{Introduction}\label{sec:intro}

Life science students are the largest population of science majors taking physics courses at colleges and universities in the United States.\cite{nsb2012} These students represent such a large fraction of the physics enrollment that many colleges offer separate introductory physics courses specifically for life sciences majors. However, the content of these physics courses is often inadequately aligned with the interests of life science students, amounting to algebra-based versions of the calculus-based courses taken by physical science and engineering majors. Recent years have seen a number of calls to improve the general state of undergraduate physics, including specific appeals to develop interdisciplinary courses that serve the needs of life sciences majors.\cite{national2012,national2013Adapting} 

Biology teaching in higher education is undergoing similar reconsiderations in light of new frontiers in biological research. Recent calls for transformation emphasize the need for students to understand biology from a systems perspective and to develop scientific practices. The focus on biological systems represents a new approach in biology education and provides an appropriate way to address biological complexity.\cite{vc2011,national2012} Key scientific practices for life science students include problem-solving, numeracy, communication, and modeling.\cite{bio20102003transforming,vc2011,national2012} This grounding in biological systems and scientific practices closely aligns with the skills needed to conduct biological research and better prepares students to be professional biologists and engaged citizens.

In the light of the changing needs of both physics and biology students\cite{vc2011} and urgent calls to develop introductory STEM courses that promote the persistence of STEM majors,\cite{pcast:2012} the physics education community is actively working to develop curricula that support the achievement and competence of life sciences majors. Several universities are developing courses in introductory physics for life science majors.\cite{meredith2013reinventing,gouvea2013framework,oshea2013f,thompson2013competency} Typically, these courses focus on modifying course content, such as adding examples grounded in biology and/or teaching laboratory techniques relevant to life science majors. However, in addition to specific content, students taking these courses must also be able to reconcile the seemingly different disciplinary practices of physics and biology.    

A quick glance at introductory textbooks in physics and biology can give the impression that these are very distinct kinds of science. Biology and physics typically differ in the types of systems that each studies, the behavior of those systems, the kinds of measurements that each makes, and the role of mathematics in each field. Because they focus on disciplinary content rather than scientific practices, textbooks often accentuate the differences between the disciplines. Physics textbooks\cite{cummings2004understanding,giancoli2008physics,knight2004physics,chabay2011matter,serway2012physics} contain dozens of equations in the first few chapters alone, whereas introductory biology textbooks\cite{reecebio,brooker2013biology,raven2013biology} contain numerous diagrams of biological systems, structures, and processes, but few mathematical equations.

While biology and physics might appear quite distinct to students, as scientific disciplines they both rely on observations and measurements to explain or to make predictions about the natural world. As a shared scientific practice, modeling is fundamental to both biology and physics. Models in these two disciplines serve to explain phenomena of the natural world; they make predictions that drive hypothesis generation and data collection, or they explain the function of an entity. While each discipline may prioritize different types of representations (e.g., diagrams, mathematical equations) for building and depicting their underlying models, these differences reflect merely alternative uses of a common modeling process. Building on this foundational link between the disciplines, we propose that teaching science courses with an overarching emphasis on scientific practices, particularly modeling, will help students achieve an integrated and coherent understanding that will allow them to drive discovery in the interdisciplinary sciences. 

The present theoretical paper makes the case that efforts to bridge biology and physics education must include an explicit emphasis on the practice of modeling.  We focus on modeling because it reconciles many of the apparent differences between these two disciplines. Using canonical examples from physics and biology, we demonstrate the ubiquity of modeling by showing how biological and physical models can be framed with a relatively small set of modeling principles, constructs, and activities common to both disciplines. Using these examples, we suggest several generalizable approaches that physics educators can take to emphasize models and modeling, especially (but not exclusively) in introductory physics courses for life science students. Our theoretical approach also suggests several areas of research in the design, implementation, and evaluation of instruction in modeling. Our aim is neither a broad treatment of the nature or meaning of scientific models and modeling, nor a narrow focus on specific content that can be ported directly into introductory courses, but rather to present a framework for instructors to think about how to blend key aspects of modeling into their own physics courses.

\section{Defining Models and Modeling}\label{sec:modeldef}

The recent NRC report on the state of undergraduate physics education {\it Adapting to a Changing World} highlights students' abilities to make models as essential to their development of a deep understanding of the physical sciences.\cite{national2013Adapting} Modeling has also been promoted more generally as a key practice in transforming science instruction.\cite{quinn2011framework,NGSS2013,vc2011} Yet, despite the occasional use of the term ``model'' in our science classrooms (e.g., Bohr model, Standard Model of particle physics, Lotka-Volterra predator-prey model), models and modeling are rarely given explicit treatment, except in particular curricula emphasizing them.\cite{wells1995modeling, etkina2007investigative,hoskinson2010build} Below, we provide operational definitions of models (products) and modeling (practices) that are broadly applicable across the sciences. By identifying the many similarities and few differences between modeling in physical and life sciences, we present a way to bridge the teaching of physics and biology through scientific practices that are fundamental to both disciplines. 

\subsection{What is a model?}

\begin{figure}[t]
\includegraphics[width=0.9\linewidth, clip, trim=0mm 22mm 32mm 42mm]{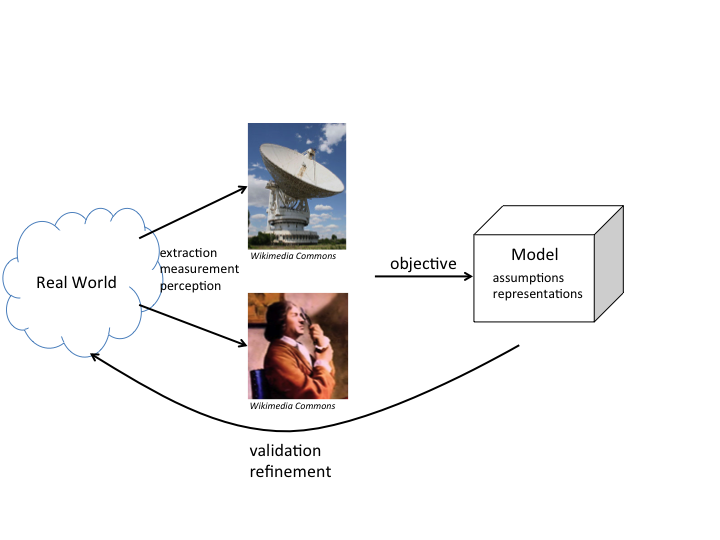}
\caption{Scientific models are simplified representations of natural systems, built to satisfy one or more objectives of explanation, prediction, or functionality. The modeling process is iterated when models are compared to real-world systems and validated or refined.}\label{fig:model}
\end{figure}

Real world phenomena consist of multiple entities (e.g. organisms, atoms, energy) interacting in complex ways. Scientists gather, organize, and analyze data and information generated by these interactions. However, not all the information in the real world can be collected and interpreted simultaneously, so we use instruments to measure phenomena and filters to extract only relevant information (Fig.\ \ref{fig:model}). A model is, fundamentally, a tool that simplifies the real world and allows us to interact with the idealized world in meaningful ways.\cite{hestenes1987toward,halloun2004modeling,Schwarz2009} Modeling is a basic human enterprise, and we build models -- simplified, conceptual constructs that help us make sense of complex and messy systems -- all the time. 
Scientific models fulfill one or more of three functional roles: they are explanatory (e.g., using structure to explain a phenomenon or observation), predictive (e.g., of a cause or effect), and/or functional (e.g., describing processes or operations).\cite{Starfield1994, Schwarz2009} The process of building and using models is termed {\it modeling}.\cite{halloun2004modeling,quinn2011framework,Schwarz2009} 

At a finer grain, models are driven by their objectives.\cite{Starfield1994} There may exist many equally valid models of the same system, and models often place greater value on the model's utility than its "correctness."\cite{Odenbaugh2005} For example, the model of an atom built to make sense of the distribution of charge (Rutherford model) may be different from the model of the same system -- the atom -- built to make sense of the hydrogen spectrum (Bohr model). Each model, therefore, intentionally focuses on a limited set of details. Thus, every model necessarily includes assumptions -- simplifying hypotheses, idealizations, or approximations.\cite{Starfield1994} In building models that necessarily simplify the real world, there is often a trade-off between simplification and robustness:  a model may explain much, but be difficult to understand; or it may be understandable to many, but explain less.\cite{levins1966strategy} Because simplifications are a necessary part of any model, the modeling process includes identifying both the relevant features and the level of detail needed to achieve the objectives of the model. 

An essential step of any modeling process occurs when we design or use representations to communicate models with other people. Model representations foster clear thought and communication about abstract ideas and their corresponding real-world systems.\cite{Schwarz2009} Within every discipline, there are standard and convenient representations that are used widely. In both physical and biological sciences, these standard representational forms include equations (with variables as symbolic notation), plain language, diagrams, pictures and animations (indicating actions or processes), and graphs. Not every equation or diagram represents a model; model representations are explanatory, predictive, or functional. In science education, the ability to represent existing and new models -- and critique the models of others -- becomes a key goal of mastering the practice of scientific modeling.\cite{Schwarz2009, Dauer2013} The ability to represent the same model system in multiple ways is also an important benchmark in students developing sophisticated conceptual understanding of physical and biological systems.\cite{McKagan2008,tsui2003genetics}

\subsection{What is modeling?}

The process of modeling includes a variety of practices, such as constructing models, making predictions, comparing predictions with measurements, refining models and explanations, and communicating model findings.\cite{Schwarz2009} A testable scientific model generates predictions that can be compared with real-world observations. Predictive models also allow scientists to explore hypothetical situations that are difficult, undesirable, or (especially in biology) unethical to initiate or observe. Predictive models also drive hypothesis generation in both physics and biology -- a key practice for both disciplines.

Once a scientist uses a model to make a prediction, the opportunity to compare predictions with measurements arises. The process of comparing observations to model predictions is dubbed {\it model testing} or {\it validating}.\cite{Starfield1994} When this process of comparison is iterated, that process is termed {\it refinement}. Model refinement is not merely adding more detail to a model. It includes adding new, relevant features and removing extraneous, unnecessary features, so that the model becomes more robust or better aligned with observations. Model validation is an iterative process centered on the development of ideas as the scope and sophistication of both models and observations increases.\cite{hestenes1987toward,halloun2004modeling,Starfield1994, Schwarz2009, Dauer2013}

Modeling also necessarily includes the iterative process of communicating scientific ideas. As sense-making tools, models are only as useful as their representations (e.g., equations, diagrams).  It is these representations that have the power to communicate an explanation, a mechanism, or a function in ways that are meaningful to others.\cite{McKagan2008, Schwarz2009, Dauer2013} Model communication is essential when evaluating particularly charged topics, such as climate change, but it is equally important for promoting student learning of non-contentious topics (e.g., models of motion or diffusion).\cite{hestenes1987toward,tsui2003genetics} 

\section{Why do biology and physics seem so different?}

Models and modeling are used throughout science, and thus models in both biology and physics have the same general features --  an idealized system, model objective, assumptions, representations, limitations, refinements, and validation measures (Tables \ref{tab:atomcell}--\ref{tab:carnotheart}). However, the unique content of each discipline means these common features of modeling may appear distinct at first glance. We describe four perceived differences in the processes of modeling between physics and biology arising from the nature of the systems that each studies.

\underline{Nature of models:} Physical models are nearly always built to make predictions, and consequently, they make extensive use of mathematical representations. Within the introductory physics course, properties like positions, velocities, forces, and energies can all be calculated as a system evolves, starting from a range of initial conditions, leading to a rich set of predictions. Most introductory biology courses utilize mathematical representations far less frequently.\cite{reecebio,raven2013biology,brooker2013biology} At first glance, the lack of mathematics may give the appearance that biological models are merely functional, while physics models are predictive. However, the objective of the biological models in introductory courses is not to make precise numerical predictions, but to explain linkages between structure and function and to drive hypothesis generation.\cite{Dauer2013} In fact, many biological models - like physical models - make sophisticated predictions that inform biological experimentation.\cite{Schwarz2009} Gregor Mendel developed the biological model of particulate inheritance -- the principle that parents pass traits to their offspring via discrete particles termed genes --  through his now-famous pea-plant breeding experiments. In turn, Mendel's predictive model drove future biologists to search for a substance within cells -- thought to be a macromolecule -- that conveyed genetic information, long before a functional or mechanistic understanding of DNA emerged.

\underline{Emphasis on different model types and objectives:} Models in physics involve entities that are reduced (e.g., fundamental particles like electrons), or simplified to the point where structure is ignored (e.g., ``rigid body'', point mass). There are exceptions, such as with condensed matter physics, where structure and function are intimately related, but these topics are not usually part of introductory curricula. Modeling in introductory physics instead emphasizes predictive or explanatory models (e.g., the trajectory of a particle, the voltage induced by a generator), rather than functional models. Since biological systems are complex, often the structure and function of systems are inextricably intertwined. Thus, many biological models in introductory biology courses begin by articulating the structure to explain the function of complex systems (e.g., a cell or an ecosystem) as scaffolds for understanding the mechanistic models that describe how those systems are built, organized, and maintained. For instance, to understand Mendel's discovery, students often learn about the structure and function of DNA, genes, and chromosomes.

\underline{Common representations may differ:} It follows that because each discipline emphasizes different model types, model representations are different as well. Physics often begins with simple idealized systems where the objective is to explain and to make predictions. When learning about mechanics, students consider simple idealized systems where the objective is to explain and to predict the motion of objects based on interactions with their environments. Introductory physics, therefore, emphasizes graphs and equations. Because biological models vary significantly across scales, and understanding structure often precedes explaining function, biological models typically utilize multiple pictorial and diagrammatic representations for a single system. For example, biology students studying chromosomes typically work with electron microscope images; schematic, idealized diagrams of chromosome pairs during cell division; and similar representations at larger and smaller scales of organization (e.g., molecular, chromosomal, cellular) when learning about genetics.\cite{reecebio}

\underline{The role of scale:} Many models in physics behave similarly across a wide range of scales. For example, classical models of an ideal gas, the motion of bullet, or the orbit of the earth around the sun all are very similar models that span an enormous range of spatial scales, from the atomic (10$^{-10}$ meters) to the astronomical (10$^{10}$ m). When spatial or temporal scales do matter in physics, it is most commonly in interpreting macroscopic bulk properties or materials (e.g., elasticity or index of refraction) in terms of a microscopic model of matter.\cite{chabay2011matter} In this sense, physics is often taught ``bottom up'':  the system of interest is reduced to fundamental basic entities and interactions that explain properties of the entire system. By contrast, biological systems are complex, hierarchical, and scale-dependent. Molecules, cells, organs, organisms, populations, and ecosystems behave differently, and new models are necessary at each different scale. Biological systems are emergent -- that is, the systems behave very differently at different spatiotemporal scales. For example, a single neuron behaves differently than a cluster of neurons, and all behaviors of the cluster cannot be predicted simply by knowing the properties of the single neuron. Properties at one scale of organization (cells) are necessary but insufficient to understand structure and function at a larger scale of organization (tissues). Because each scale range requires development of some new models, the breadth of an introductory biology course is large.

\section{Examples of models and modeling physics and biology}

Using the frameworks of a model and knowing how scientists engage in modeling, we consider pairs of models -- one each from physics and biology -- that illuminate the common uses of models in each discipline. We also highlight the elements of models and modeling that are unique to each discipline. The first pair of models, the atom and the cell, represent models developed primarily to explain the structure of the building blocks of matter and life. The second example, on the motion of cars and bacterial colonies, illustrates a pair of predictive models from physics and biology, respectively. The third pair of models represent functional models that deconstruct the processes by which thermal energy is transformed into mechanical energy, in the Carnot cycle and the animal heart. We present these models as examples to suggest general approaches that physics instructors can employ to help biology students develop the practices of modeling.

\subsection{Structure Function models: Cells and Atoms}\label{sec:atomcell}

\begin{table*}
\begin{tabular}{|p{0.16\linewidth}|p{0.27\linewidth}|p{0.27\linewidth}|p{0.27\linewidth}|}\hline
{\bf Real World} & \multicolumn{2}{|l|}{Atomic systems} & Living cells \\\hline
{\bf Idealized System} & Rutherford atom & Bohr atom & Typical eukaryotic cell \\\hline
{\bf Model objective} & Explain observations of a dense, central charge & Describe how nested electrons produced the hydrogen spectrum & Describe the composition of the smallest units of life\\\hline
{\bf Assumptions} & Atoms have a massive, fixed, positively-charged nucleus; Scattering results from interactions of $\alpha$-particle and positively charged nucleus; Other sources of $\alpha$-particle scatter can be ignored & Stationary electrons emit no radiation; Atoms have a discrete radius, and electrons occur in discrete, quantifiable orbits; Electrons moving between discrete orbits must absorb or emit energy (in photons) to complete the transition & All living things are composed of one or more cells; All cells come from other cells; all cells contain the hereditary information required to produce more of their own type; Life functions occur within (and among) cells\\\hline
{\bf Representation} & \begin{center}\includegraphics[clip, trim=70mm 50mm 75mm 40mm, width=0.5\linewidth]{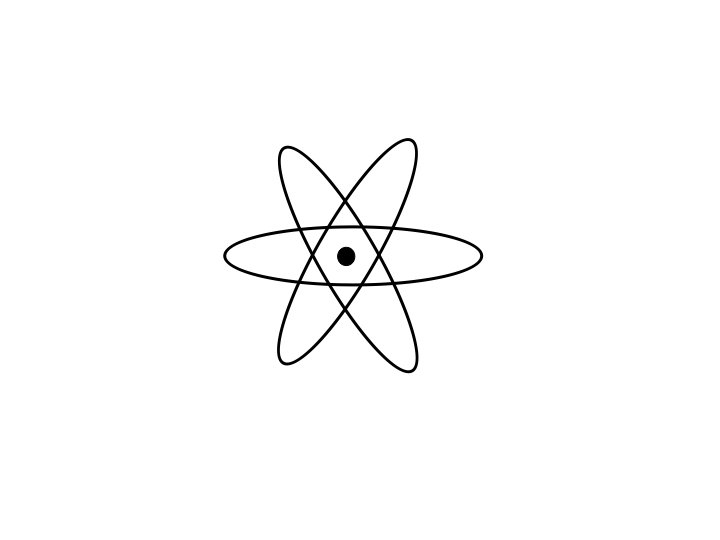}\end{center} & \begin{center}\includegraphics[clip, trim=70mm 50mm 75mm 40mm, width=0.5\linewidth]{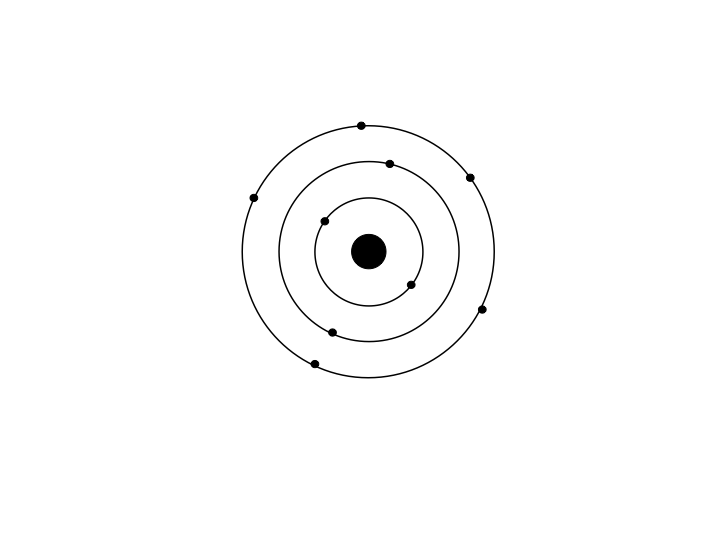}\end{center} & \begin{center}\includegraphics{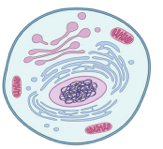}\end{center} \\\hline
{\bf Limitations} & Electrons in diffuse orbits lose energy while orbiting and collapse in an unstable atom & Does not accurately represent larger atoms with multiple electrons in multiple energy states & May not represent cells with different arrangements of organelles\\\hline
{\bf Refinement} & Improved ``plum pudding'' model by proposing a massive nucleus and separate regions of charge & Improved Rutherford model by discretizing electron orbits and corresponding transition energies & Huge diversity of cells; structure/function relationship makes it impossible to represent all cells with a single generic cell\\\hline
{\bf Validation} & Observation of dense, positive atomic core (Geiger, Marsden)	& Observation of the spectral lines of hydrogen (Lyman, Balmer, others) &	Observation of cells and cellular structure (Hooke, Pasteur)\\\hline
\end{tabular}\caption{Examples and characteristics of {\it explanatory} models in physics (of the atom) and biology (the cell).}\label{tab:atomcell}
\end{table*}

Cells and atoms are canonical structures that serve as quintessential examples of the modeling process and its value in the sciences.\cite{cummings2004understanding,giancoli2008physics,knight2004physics,chabay2011matter,serway2012physics,reecebio} It is common in chemistry and in modern physics courses to lead students successively through more sophisticated models of the atom. Often, instructors begin with Thomson's ``plum-pudding'' model and progress through Rutherford's ``planetary'' model, Bohr's ``solar system'' model, and Schrodinger's ``cloud'' model. For simplicity, we have referenced the Rutherford and Bohr models to illustrate the importance of model refinement in the scientific modeling process (Table \ref{tab:atomcell}). Rutherford's objectyive was to explain the scattering of alpha particles in terms of the distribution of charged particles in the atom. Rutherford's model, though, was unable to explain the emission and absorption spectra of hydrogen, a phenomenon explained by the Bohr model. However, Bohr's model cannot describe the structure and dynamics of the atom, the later objective of Schrodinger's model. Each model captures different features and behaviors of the atom, and each makes different assumptions about an atom's behavior. In each instance after Thomson's, prior models were tested and refined. Inaccuracies -- such as non-discrete electron orbitals -- were stripped away as each model was tested against observation and found unable to explain certain phenomena. By focusing on the practice of modeling as an iterative, dynamic process, physics instructors can demonstrate that sound science is not about finding the one ``true'' model of the atom, but showing that depending on the available observations and level of detail required, many models may be useful scientific models, although they may have limitations. An explicit discussion of scientific models and modeling highlights the features that make a model useful for a given purpose, and leads to a discussion of what gives one model greater utility or parsimony than another.

As the atom is the building block of matter, the cell is the building block of life. Cells are frequently modeled in biology curricula, and just as with physical models of the atom, the choices made in the modeling process depend on the objectives of the model. Here, we focus on a model of a cell, with the purpose of explaining the composition of an idealized cell as the basic unit of life. A key assumption of early cell models (since tested and found valid) is that cells are irreducibly alive; all functions of life, then, occur within cells. Models of idealized cells may be represented by ``cartoon'' diagrams -- that is, highly simplified, roughly scaled drawings usually with false-colored structures. As with models of the atom, successive models of cells have been tested, validated, and refined, with new features added and extraneous features stripped as new observations invalidated previous assumptions. Some of the first models of the cell were van Leeuwenhoek's seventeenth-century drawings of ``animalcules'' (microorganisms). Startling in their detail, these were the first representations of cells viewed through a handheld, handmade, first-generation microscope. However, van Leeuwenhoek's model of a sperm cell, in which he thought he observed a tiny homunculus (a fully-formed individual), was subsequently revised as more powerful microscopes afforded more accurate images.

This pair of examples is meant to highlight the essential steps of model validation and refinement. Physics instructors can reinforce this part of the modeling process with all their students by choosing and designing problems and cases that necessarily build on previous models, so that students develop experience in the sustained and iterative process of model-building as an essential scientific practice.

\subsection{Predictive models: Motion of a car and of a bacterial colony}\label{sec:car}

The topics of motion and forces make up a large fraction of introductory physics content. Models of motion are often predictive. These models refer to some particular real-world phenomenon (e.g., the motion of a car) for which the objective is to predict: a trajectory, velocity, acceleration, or some other kinematic property. Initial versions of these models used in instruction do not account for the mechanisms causing this motion (i.e., forces). In predictive models such as models of motion, equations and graphs are common forms of representation. With a foundation of kinematic models, introductory physics courses typically proceed to dynamics, where interacting forces are now included. Gravitational, spring-based, and frictional forces serve as new variables within the model and allow motion prediction across a wide range of circumstances, depending on the particular force. 

In biology, the directed motion of groups of cells, termed taxis, is important for cells seeking food or avoiding harm. Chemotaxis allows bacteria to find glucose on a culture plate or avoid antibiotic toxins; it also allows sperm cells to locate oocytes prior to fertilization. Biologists use models to predict the same types of parameters that physicists do:  a mean path (akin to center of mass motion), a trajectory of motion, and a rate of motion. When modeling chemotaxis, biologists typically assume that the organisms choose an optimal path. Biologist frequently consider populations, or groups of similar-type individuals (whether cells or organisms). Many predictive models in biology are therefore probabilistic, emerging from the collective behavior of many individuals -- an important difference between predictive models in physics and biology. As in physics, predictive models in biology are often represented using equations and graphs. Although chemotaxis models provide quantitative predictions of rates and trajectories for a particular chemotactic agent acting on a group of cells, the effect of a new chemotactic agent cannot be predicted.

This pair of predictive models (Table \ref{tab:carcell} in Appendix \ref{sec:other}) shows that physics instructors who wish to emphasize the development of modeling skills can focus on aligning models with their representations. Predictive models tend to use equations and graphs because scientific predictions are often quantitative. Physics instructors can also choose problems and cases that illuminate the principles of predicting a population-level, or probabilistic, outcome. Furthermore, since many scientific models have more than one objective (e.g., explanatory and predictive), such models may employ multiple representations, such as equations and diagrams. Life sciences students may need frequent practice in negotiating among the representations -- for example, translating equations into graphs.

\subsection{Models of processes: The Carnot cycle and the heart}\label{sec:carnot}

\begin{figure}
        \centering
        \begin{subfigure}[b]{0.9\linewidth}
                \centering
                \includegraphics[clip,trim=10mm 15mm 10mm 10mm,width=\textwidth]{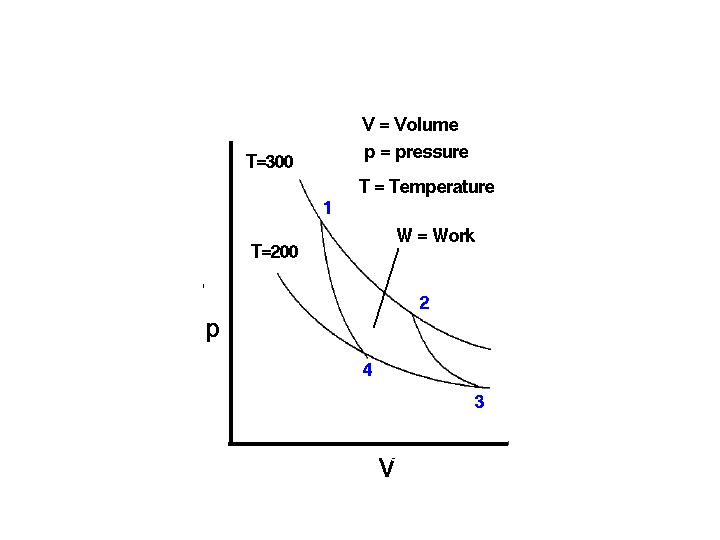}
                \caption{P-V diagram of an idealized Carnot engine}
                \label{fig:carnot}
        \end{subfigure}
        ~
        \begin{subfigure}[b]{0.9\linewidth}
                \centering
                \includegraphics[clip,trim=0mm 60mm 50mm 0mm,width=\textwidth]{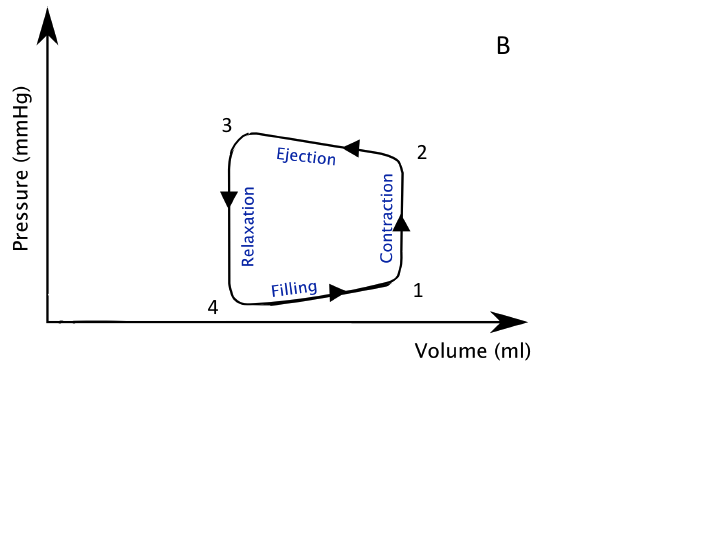}
                \caption{P-V diagram of the heart}
                \label{fig:heart}
        \end{subfigure}
        \caption{Functional models often evoke diagrammatic representations, such as these P-V diagrams of a Carnot engine (a) and of an idealized animal heart (b).}\label{fig:pv}
\end{figure}

Physicists exploring introductory biology textbooks may be surprised at just how many of the diagrams portray cycles and processes: life cycles, cell cycles, nutrient cycles, developmental processes, cycles of respiration and photosynthesis. The principles of matter and energy conservation underlie many of these diagrams portraying complicated biological systems. The connection between energy and entropy is critically important for developing systems thinking in introductory biology courses.\cite{vc2011,rice2013}

The process by which usable work is cyclically extracted from a heat engine is a fundamental example reflecting the principles of energy conservation and entropy. A physical model of this process, the Carnot engine, is typically represented in a diagram with four stages. The objective of this model is to describe the process by which thermal energy is converted into mechanical energy as energy flows from one reservoir to another through the engine (i.e., the Carnot cycle). For simplicity, we usually assume that the system containing a hot and a cold reservoir is closed and can be completely described by energy conservation and the second law of thermodynamics. Hence, there is no description of the time over which this process occurs. This process is commonly represented graphically in P-V diagrams (Fig.\ \ref{fig:pv}). 

A Carnot cycle of a gaseous system, such as one describing an internal-combustion engine, begins with the isothermal compression of a gas (1$\rightarrow$2). The gas is then further compressed adiabatically (2$\rightarrow$3). At this point, in the real system, the gas is ignited and expands the volume (ignition, point 3). In the Carnot cycle, this expansion is isothermal (3$\rightarrow$4). The gas then expands adiabatically (4$\rightarrow$1) before the process is repeated. The total work done by an internal combustion engine is the area of the space bounded by the curves in the P-V diagram (Fig. \ref{fig:carnot}). Because this model is primarily functional (process), it does not predict the rate of work by the engine. 

A biological example of the heat engine is the animal heart. Animal hearts use energy, supplied from cellular respiration, to do the mechanical work required to move blood throughout the animal's body. Consider a left ventricle of a four-chambered human heart (the left ventricle sends oxygenated blood into the aorta and throughout the body). At the end-diastolic point (Fig.\ \ref{fig:heart}, 1), the left ventricle begins to contract, until the pressure inside it exceeds the pressure in the aorta (2), and the blood inside the ventricle is ejected. Blood volume in the ventricle decreases and pressure falls (3), triggering the opening of the mitral valve that initiates ventricular filling (4). As long as the animal can consume food to fuel respiration, the cycle repeats indefinitely throughout the organism's lifetime.

This pair of functional models (Table \ref{tab:carnotheart} in Appendix \ref{sec:other}) demonstrates how models from different systems can be used to convey overarching thermodynamic principles, which are found throughout biological systems (for example, photosynthesis and respiration can be characterized by thermodynamic processes, as can some homeostatic processes). The Carnot process invites physics teachers to emphasize the conservation of matter and energy -- a core concept in biology as well as physics.\cite{vc2011}  Students have distinct difficulties tracing matter and energy through biological systems, believing, for example, that matter and energy can be ``used up'' or that matter can change into energy.\cite{rice2013} Physics teachers can emphasize this and other models of thermodynamic processes that consider matter and energy as distinct, but mutually necessary elements of a real-world system and models of the system.

The Carnot model specifically, and thermodynamic models generally, also illustrate the importance of navigating among the multiple representations of many scientific models. While the Carnot engine and ventricular-function models have clear functional objectives, they are also used to predict how much total work might be done, or the total blood volume able to be moved by a heart. Such predictive features of these models invite mathematical representations. Biology students do not see many mathematical models in their introductory courses, so physics teachers can serve students well by emphasizing mathematical representations when they are engaged with predictive models.

\section{Implications}

In this paper, we have claimed that modeling is a ubiquitous scientific practice, and that a small set of operational elements and principles govern scientific models and modeling in the physical and life sciences. Hence, one powerful way to support interdisciplinary science teaching is to focus instruction on the practice of modeling. Furthermore, these claims suggest possible research questions that explore the design, implementation, and evaluation of model-based instruction.

Science is the practice of modeling, and our examples demonstrate that modeling is fundamentally a shared practice of physics and biology. Models are what we use to make sense of our world; they are tools for understanding natural phenomena. Life science majors would be well-served in physics courses where an explicit emphasis is placed on the process of constructing and validating physical models. And yet, most introductory physics courses under-emphasize the concept of a model and the process of modeling. This deficiency is unfortunate, because in their future course work, students will use biological models extensively and engage in the modeling process. Additionally, the modeling process closely approximates the types of skills that professional biologists will employ throughout their careers. The centrality of modeling as a scientific practice naturally prompts research questions related to the similarities and differences between models and modeling in physical and life science instruction. For instance, how does learning modeling differ between physics and life science courses? 

While physics and biology appear to have rather distinct disciplinary practices, we argue that an emphasis on modeling can serve to reconcile these two disciplines at their most fundamental level.  As a shared practice within the sciences, modeling provides a common framework from which specific disciplinary differences can be compared and elaborated. Differences in the selection of model types result from differences inherent to the systems that each discipline studies, the types of models built (e.g. descriptive, predictive, or functional), and common representations (diagrams, graphs, equations, narratives). Unpacking and engaging in the modeling process provides ample opportunities for both instructors and students to describe how and why models in each discipline are developed and used. Specific models from the discipline of biophysics provide an additional means of bridging these two disciplines. Given the emphases on different model types between biology and physics, an important question arises: To what extent are students able to transfer knowledge of the modeling process across biology and physics, and more broadly, among disciplines that emphasize modeling as a practice?

The modeling practice is applicable to every scientific discipline and uses only a small set of core modeling principles. As a result, modeling provides a common language for cross-disciplinary dialogue. Models in biology and physics are defined by one or more objectives; they explain phenomena, they make predictions, and / or they describe functions and processes. All models make simplifying assumptions, employ representations that are aligned with model objectives, and are validated and refined through iterative interaction between experimental observations and model outputs. While specific representations may differ between disciplines, these differences can be understood through their alignment to model objectives.  Model representations are essential forms of scientific communication, and an interesting research question relevant across disciplines is:  Do certain representations better support students' sense-making about the model or the real world? Closely related is the question, how do multiple representations of one model support students' model-building and sense-making skills?

The most common approaches to science curricular reform focus on changes in course content or context. Indeed, a number of teams are working on specific content changes to their introductory courses.\cite{meredith2013reinventing,gouvea2013framework,oshea2013f,thompson2013competency} While these thoughtful approaches are encouraging, it is unlikely that these wholesale course revisions will be adopted broadly, because content-specific changes to the introductory physics courses will ultimately be informed by local conditions. We have provided a complementary view to transforming undergraduate science courses by illustrating how physics and biology are united in their underlying use of scientific models, and by describing how this practice can be leveraged to bridge the teaching of physics and biology. Our view complements that of curriculum developers by providing a theoretical scaffold on which to center individual and community efforts. Our work will be further enhanced as curriculum developers and education researchers develop deeper understandings of the mechanisms and impacts of model-based instruction. 

\begin{acknowledgments}

The authors wish to thank D. Ebert-May for reading an early draft of this manuscript, and J. Burk, S. Chivukula, S. Douglas, B. O'Shea, T. Long, J. Middlemas-Maher, and two anonymous reviewers for comments that improved the manuscript. This work was supported by the University of Colorado's Science Education Initiative and the National Science Foundation's Division of Undergraduate Education (DUE-1043028).

\end{acknowledgments}

\appendix
\section{Other model types}\label{sec:other}

The following tables apply the framework developed in Sec.\ \ref{sec:modeldef} to the models presented in Secs.\ \ref{sec:car} and \ref{sec:carnot}.

\begin{table*}[b]
\begin{tabular}{|p{0.16\linewidth}|p{0.405\linewidth}|p{0.405\linewidth}|}\hline
{\bf Real World} & Car & Bacterial colony (10$^8$ members) \\\hline
{\bf Idealized System} & Point particle & Small collection of cells \\\hline
{\bf Model objective} & Predict properties of motion (rate, trajectory) & Predict the rate or trajectory of movement\\\hline
{\bf Assumptions} & Forces (gravity, spring, friction) are initially ignored&All cells of a similar type respond to cues in the same ways\\\hline
{\bf Representation} & Often, equations and graphs&Equations, graphs \\\hline
{\bf Limitations} & Does not explain {\it how} the car moves, or how an internal-combustion engine works.
	& Typically does not explain how cells move; Does not explain how chemotactic substances work (e.g. signal transduction); Although the group's motion can be characterized probabilistically, the model cannot predict how a particular individual will move
\\\hline
{\bf Refinement} & Dynamical systems can be understood by relaxing the assumption that forces are unimportant.&Chemotactic models that account for differences between cell types and mechanisms of cell motility may enhance predictive power.\\\hline
{\bf Validation} & Observation of motion & Observation of collective motion \\\hline
\end{tabular}\caption{Examples and characteristics of {\it predictive} models in physics (a body in motion) and biology (chemotaxis).}\label{tab:carcell}
\end{table*}

\begin{table*}[b]
\begin{tabular}{|p{0.16\linewidth}|p{0.405\linewidth}|p{0.405\linewidth}|}\hline
{\bf Real world} & Combustion engine & Animal heart \\\hline
{\bf Idealized System} & Carnot engine & Cardiac heat engine \\\hline
{\bf Model objective} & Describes how heat (thermal energy) is converted into mechanical energy to do work&Describes how a heart converts heat into the mechanical energy required to pump blood\\\hline
{\bf Assumptions} & System is quasi-closed; Reservoirs are infinitely large; All the processes are reversible&Variance in heat or blood volumes can be ignored
\\\hline
{\bf Representation} & P-V diagrams of an engine piston	& P-V diagrams of a heart chamber \\\hline
{\bf Limitations} & Does not predict the rate or total production of work; Is highly idealized &Describes only the process of ventricular rhythm, not its rate or total blood volume moved\\\hline
{\bf Refinement} & Otto cycle can describe functioning engine;
Can add detail of two-stroke or four-stroke engine & Can be modified to describe two-, three-, or four-chambered hearts; Can add details of organisms' morphology
\\\hline
{\bf Validation} & Comparison to idealized engine & Comparison to cardiac function in organisms with single-chambered hearts \\\hline
\end{tabular}\caption{Examples and characteristics of {\it functional} models in physics (the heat engine) and biology (cardiac function)}\label{tab:carnotheart}
\end{table*}



\end{document}